\PassOptionsToPackage{hyphens}{url}
\documentclass[%
 aip,
 amsmath,amssymb,
 reprint,%
]{revtex4-1}

\usepackage{graphicx}
\usepackage{dcolumn}
\usepackage{bm}

\usepackage[utf8]{inputenc}
\usepackage[T1]{fontenc}
\usepackage{mathptmx,comment}
\usepackage{amsmath, amssymb, mathtools, bm, etoolbox, comment}
\usepackage{setspace}
\usepackage{mathrsfs}
\usepackage{stmaryrd}
\usepackage{amsthm}
\usepackage{amsfonts}
\usepackage{graphicx}
\usepackage[colorinlistoftodos]{todonotes}
\usepackage{tikz-cd}   
\usepackage{booktabs}       
\usepackage{nicefrac}       
\usepackage{microtype}      
\usepackage[utf8]{inputenc} 
\usepackage[T1]{fontenc}    
\usepackage{hyperref}       
\usepackage{url}            
\usepackage{lingmacros}
\usepackage{tree-dvips} 
\usepackage{nccmath}
\usepackage{bbm}

\usepackage[utf8]{inputenc} 
\usepackage[noend]{algpseudocode}
\usepackage{algorithm}
\usepackage{caption}
\usepackage{subcaption}
\usepackage{color}
\usepackage{enumitem}

\begin{document}

\preprint{AIP/123-QED} 

\title[Cluster-based dual evolution for multivariate time series: analyzing COVID-19]{
Cluster-based dual evolution for multivariate time series: analyzing COVID-19}

\author{Nick James}
\affiliation{ 
School of Mathematics and Statistics, University of Sydney, NSW, 2006, Australia
}%

\author{Max Menzies}
\email{max.menzies@alumni.harvard.edu}
\affiliation{%
Yau Mathematical Sciences Center, Tsinghua University, Beijing, 100084, China 
}%

\date{7 May 2020}
\begin{abstract}

This paper proposes a cluster-based method to analyze the evolution of multivariate time series and applies this to the COVID-19 pandemic. On each day, we partition countries into clusters according to both their case and death counts. The total number of clusters and individual countries' cluster memberships are algorithmically determined. We study the change in both quantities over time, demonstrating a close similarity in the evolution of cases and deaths. The changing number of clusters of the case counts precedes that of the death counts by 32 days. On the other hand, there is an optimal offset of 16 days with respect to the greatest consistency between cluster groupings, determined by a new method of comparing affinity matrices. With this offset in mind, we identify anomalous countries in the progression from COVID-19 cases to deaths. This analysis can aid in highlighting the most and least significant public policies in minimizing a country's COVID-19 mortality rate.

\end{abstract}

\maketitle

\begin{quotation}

COVID-19 has resulted in a global pandemic with severe human, social and economic costs. In order to manage the economic ramifications of prioritizing citizen safety, policymakers have sought a multi-level approach involving social distancing, business closures and movement restrictions. For this purpose, a careful identification of the most and least successful countries at responding to the spread of COVID-19 is of great relevance. This paper meets such demand by developing a new method to analyze \emph{multivariate time series}, in which the variables are the cumulative case and death counts of each country on each day. We have three goals: first, we analyze the case and death counts on a country by country basis; second, we analyze the two multivariate time series in conjunction to elucidate their similarity further; third, we determine anomalous countries relative to cases and deaths.

\end{quotation}

\section{\label{sec:level1}Introduction
}
\label{Introduction}

Understanding the trajectories of COVID-19 case and death counts assists governments in anticipating and responding to the impact of the pandemic. As the disease spreads, the timely identification of anomalous countries, both successful and unsuccessful, provides opportunities to determine effective response strategies. This analysis can be difficult as death counts naturally lag behind case counts.

This paper builds on a long literature of \emph{multivariate time series analysis}, developing a new mathematical method and a more extensive analysis of COVID-19 dynamics than previously performed. Existing methods of time series analysis include parametric models \cite{Hethcote2000} such as exponential  \cite{Chowell2016} or power-law models, \cite{Vazquez2006} and nonparametric methods such as distance analysis, \cite{Moeckel1997} distance correlation\cite{Szkely2007,Mendes2018,Mendes2019} and network models. \cite{Shang2020} Both parametric and nonparametric methods have been used to model COVID-19. \cite{Manchein2020, Machado2020} 

\emph{Cluster analysis} is another common statistical method with successful applications to COVID-19 and more broadly, epidemiology. Designed to group data points according to similarity, cluster analysis has been used to study non-communicable diseases,  \cite{Cassetti2008,Alashwal2019} infectious diseases,\cite{Xiao2016, McCloskey2017} and epidemic outbreaks such as Ebola, \cite{Muradi2015}  SARS \cite{Rizzi2010} and COVID-19. \cite{Machado2020}  Clustering algorithms are highly varied - common examples are K-means \cite{Lloyd1982} and spectral clustering,\cite{Luxburg2007} which partition elements into discrete sets, and hierarchical clustering,\cite{Ward1963, Szekely2005} which does not specify a precise number of clusters. In this paper we will use hierarchical clustering,\cite{Ward1963, Szekely2005} K-means \cite{Lloyd1982} and its optimal one-dimensional variant Ckmeans.1d.dp. \cite{Wang2011} K-means and Ckmeans.1d.dp require an initial choice of the number of clusters $k$. We draw upon several methods to address the subtle question of how to select this $k$. 
The goal of this paper is to use a dynamic and smoothed implementation of cluster analysis to study the worldwide spread of COVID-19, track the relationships between different countries' case and death counts, and make inferences regarding the most successful strategies in managing the progression from cases to deaths.


This paper is structured as follows: in each of the proceeding three sections, we introduce portions of our methodology and present our results. Section \ref{sec: individual} investigates the multivariate time series of cases and deaths individually. Section \ref{LearningSystemEvolutionOffset} analyzes the two time series in conjunction, determining suitable offsets for the number of clusters and the cluster memberships. Section \ref{AnomalyAnalysis} determines anomalous countries with respect to cases and deaths. Section \ref{sec: conclusion} summarizes the results and the new findings regarding COVID-19. 


\section{Individual analysis of COVID-19 cases and deaths}
\label{sec: individual}

\subsection{Time-varying cluster analysis methodology}
\label{TimeVaryingClusterAnalysis}

The most general setup of our methodology is as follows: let $x_i^{(t)}$  be a multivariate time series over an interval of length $T$, for $i=1,\dots,n$ and $t=1,\dots,T$, with each  $x_i^{(t)}$ belonging to a common normed space $\mathfrak{X}$. Slightly different procedures apply if $\mathfrak{X}$ is one-dimensional, namely $\mathbb{R}$, or higher-dimensional.

In this paper, the two multivariate time series we present are the cumulative daily counts of cases and deaths on a country by country basis. We order the countries by alphabetical order and denote these counts by $x_i^{(t)}, y_i^{(t)} \in \mathbb{R}$ respectively. We choose cumulative counts to best analyze the evolution of the disease over time. Our data spans 31/12/2019 to 30/4/2020, a period of $T=122$ days across $n=208$ countries. 

Given the exponential nature of the data, we choose a logarithmic difference as our metric. First, we do the following data preprocessing: any entry in the data that is empty or 0 - before any cases are detected - we replace with a 1, so that the log of that number is defined. Then we define a distance on case and death counts by $d(x,y)=|\log(x) - \log(y)|$. Effectively, this pulls back the standard metric on $\mathbb{R}$ under the homeomorphism $\log: \mathbb{R}^+ \to \mathbb{R}$ and makes the positive real numbers a one-dimensional normed space.

The goal is to partition the counts $x_1^{(t)},\dots,x_n^{(t)}$ into a certain number of clusters at each time $t$. We wish to carefully choose the number of clusters in such a way that provides us meaningful inference on how the data changes. A wildly varying number of clusters would obscure inference on individual countries' cluster memberships changing with time. Thus, we combine several methods of choosing this number to reduce the bias in our estimator and perform additional exponential smoothing to yield a suitably changing number with time. In our experiments, we use six methods outlined in Appendix \ref{existingclustertheory}. These have been chosen after experimentation and consultation with the literature, but our method is flexible and could use any combination of methods. Given cluster numbers $k^{(t)}_1,\dots,k^{(t)}_6$ offered by these methods, we compute the average $k^{(t)}_{av}=\frac{1}{6}\sum_{j=1}^6 k^{(t)}_j$. This is not necessarily an integer; we do not compute clusters directly with this value. 

In our implementation, this average value $k^{(t)}_{av}$ exhibits itself as approximately locally stationary. Thus, we apply exponential smoothing to $k^{(t)}_{av}$ to produce a smoothed integer value $\hat{k}^{(t)}$. We use this value $\hat{k}^{(t)}$ at each $t$ to obtain a clustering at that time. As the daily case and death data is one-dimensional, the most appropriate clustering method is the optimal implementation of K-means specific to one-dimensional data, Ckmeans.1d.dp. \cite{Wang2011}. We implement this algorithm to group daily counts into $\hat{k}^{(t)}$ clusters and sort the clusters according to the ordering on $\mathbb{R}$.

Similar experiments can also be performed for higher-dimensional data. Analyzing 3-day rolling counts of cases and deaths $\mathbf{\tilde{x}}_i^{(t)}, \mathbf{\tilde{y}}_i^{(t)} \in \mathbb{R}^3$ requires the use of standard K-means clustering. These yield similar results to the daily analysis and can be seen in Appendix \ref{3daysection}.

\subsection{Matrix analysis of multivariate time series}
\label{MatrixAnalysis}

We record the results of this analysis in several sequences of matrices. Having performed the data preprocessing described above, first let $D^{(t)}$ be the $n \times n$ matrix of (logarithmic) distances between counts $x_i^{(t)}$ at time $t$, that is, $D^{(t)}_{ij}=|\log(x_i^{(t)}) - \log(x_j^{(t)})|$. Next, let $\text{Aff}^{(t)}$ and $G^{(t)}$ be two different $n \times n$ \emph{affinity matrices} defined as follows:
\begin{align}
\text{Aff}^{(t)}_{ij}&= 1 - \frac{D^{(t)}_{ij}}{\max D^{(t)}}, \\
G^{(t)}_{ij}&=\exp \bigg( \frac{-m^2\big(D^{(t)}_{ij}\big)^2}{2(\max D^{(t)})^2} \bigg)
\label{affinity equations}
\end{align}
We term $\text{Aff}^{(t)}$ and $G^{(t)}$ \emph{standard and Gaussian affinity matrices} respectively. These definitions are motivated by standard constructions, but we appropriately normalize $G$ for subsequent analysis. We vary $m=1,2,3$ in experiments so the matrix entries mimic Gaussian spreads over $1,2,3$ standard deviations respectively. Then, let $\text{Adj}^{(t)}$ be an $n \times n$ \emph{adjacency matrix} defined as follows:
\begin{equation*}
    \text{Adj}^{(t)}_{ij} =
    \begin{cases}
      1 &  x_i^{(t)} \text{ and } x_j^{(t)} \text{ are in the same cluster}\\
      0, & \text{ else}
    \end{cases}
\end{equation*}
\begin{figure*}
    \centering
    \begin{subfigure}[b]{0.45\textwidth}
        \includegraphics[width=\textwidth]{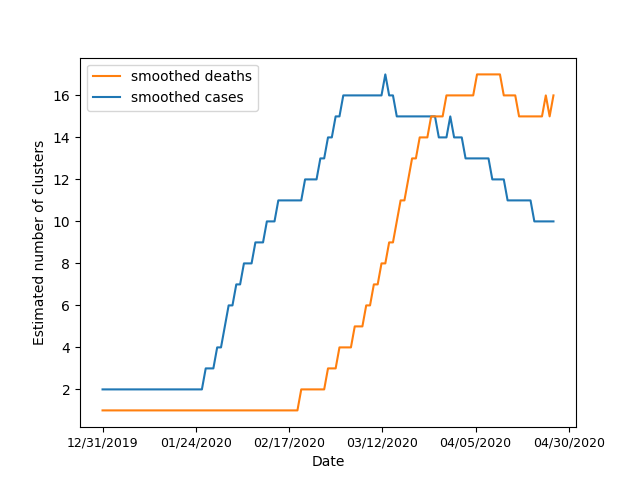}
        \caption{ }
        \label{fig:Offset_cases_deaths}
    \end{subfigure}
    \begin{subfigure}[b]{0.45\textwidth}
        \includegraphics[width=\textwidth]{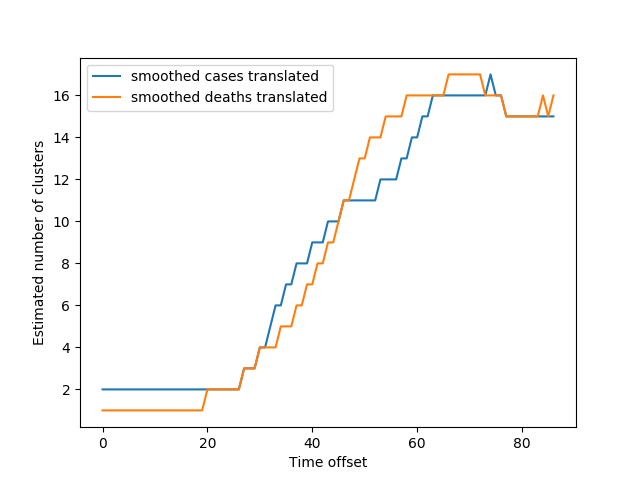}
        \caption{}
        \label{fig:Translation_k}
    \end{subfigure}
    \caption{Smoothed number of clusters $\hat{k}^{(t)}$ as a function of time, defined in Section \ref{TimeVaryingClusterAnalysis}. In (a), the blue and orange curves track the number of clusters for cases and deaths, respectively, from 12/31/2019 to 4/30/2020. In (b), the curves are shown after translation by the optimal \emph{series evolution offset}, defined in Section \ref{LearningSystemEvolutionOffset}, computed to be $\delta=32$. There is a strong similarity between the two curves up to this offset: both peak at 17 clusters before declining, suggesting reduced spread in the data.}
    \label{fig:movingk}
    \end{figure*}
    
\begin{figure*}
    \centering
    \begin{subfigure}[b]{0.45\textwidth}
        \includegraphics[width=\textwidth]{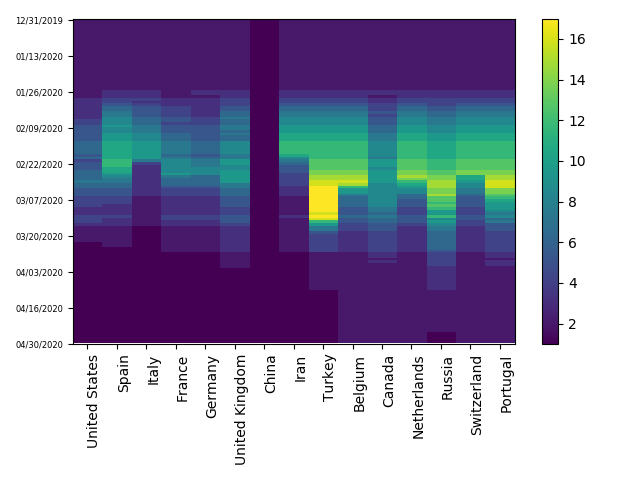}
        \caption{}
    \label{fig:CasesHeatmap}
    \end{subfigure}
    \begin{subfigure}[b]{0.45\textwidth}
        \includegraphics[width=\textwidth]{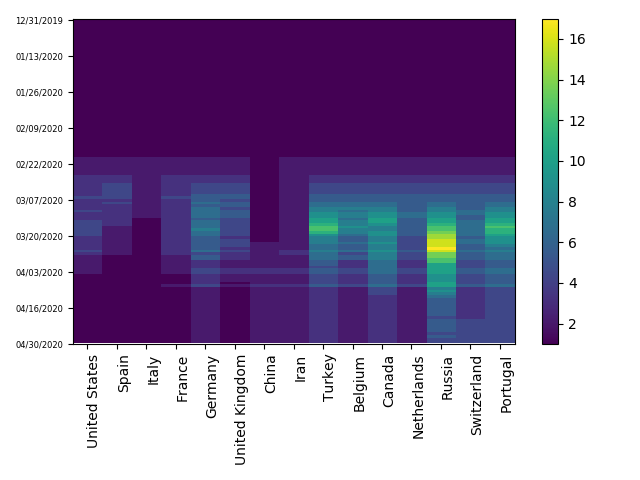}
        \caption{}
    \label{fig:DeathsHeatmap}
    \end{subfigure}
    \caption{Heat maps track the changing cluster membership of the fifteen most severely impacted countries with respect to their counts of (a) cases and (b) deaths, respectively. Cluster membership, determined by Ckmeans.1d.dp, depicts COVID-19 severity relative to the rest of the world. Clusters are ordered with 1 being the worst impacted at any time. Darker and lighter colors correspond to smaller and greater numbered cluster labels and represent worse and less affected clusters, respectively.}
    \label{fig:heatmaps}
\end{figure*}

\begin{figure*}
    \centering
    \begin{subfigure}[b]{0.85\textwidth}
        \includegraphics[width=\textwidth]{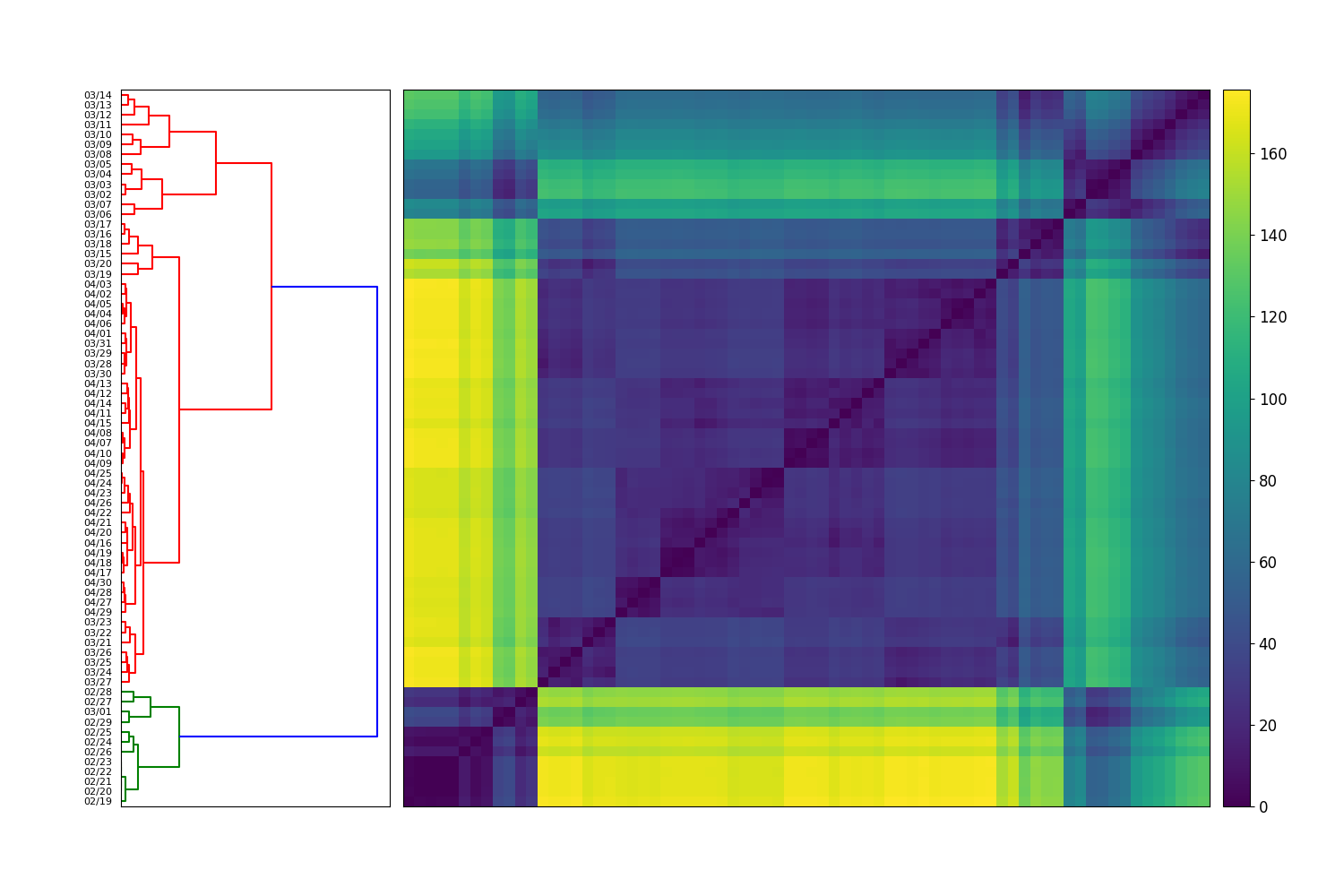}
        \caption{}
        \label{fig:Cases_daily_dendrogram}
    \end{subfigure}
    \begin{subfigure}[b]{0.85\textwidth}
        \includegraphics[width=\textwidth]{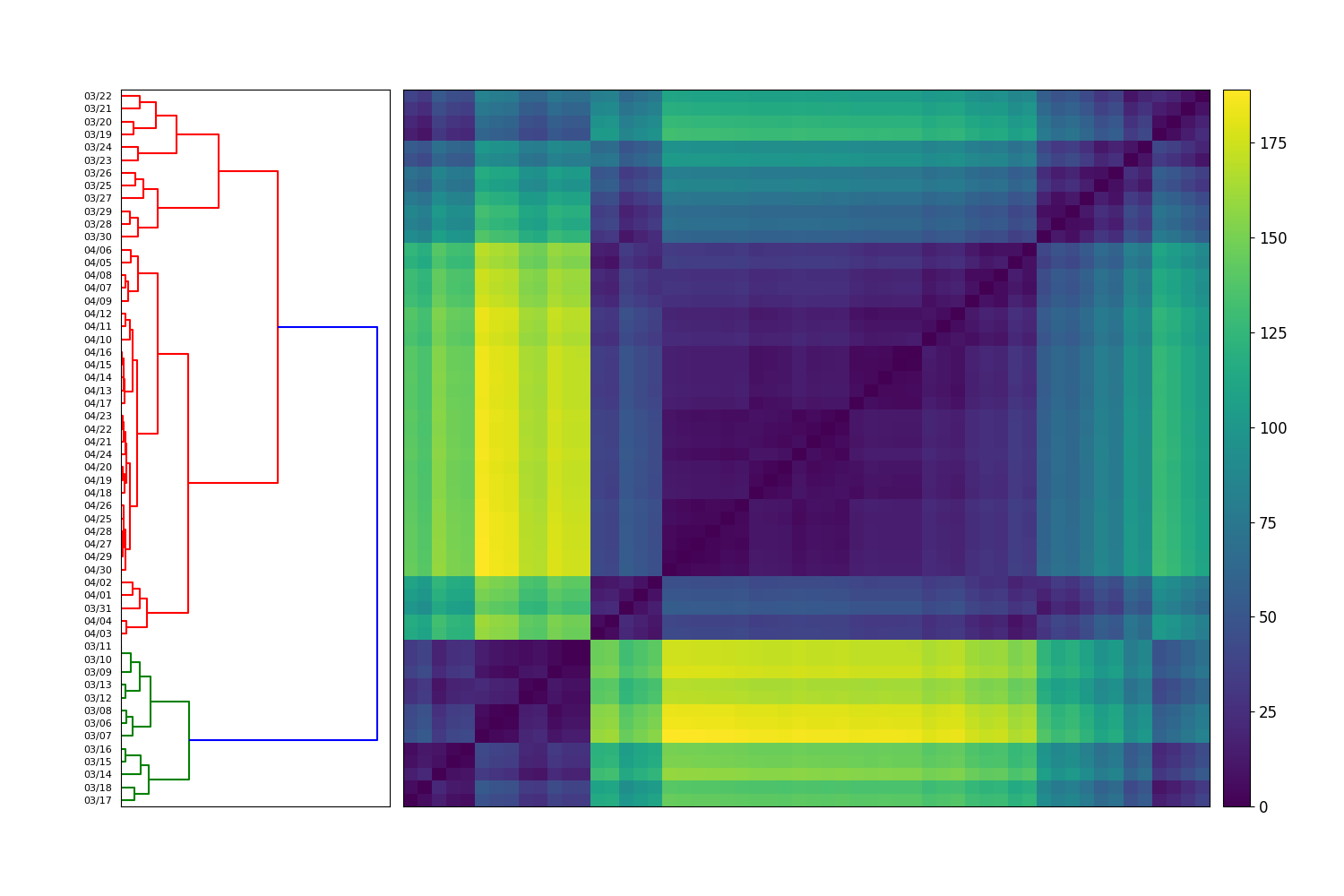}
        \caption{}
        \label{fig:Deaths_daily_dendrogram}
    \end{subfigure}
    \caption{\emph{Cluster evolution dendrograms}, defined in Section\ref{MatrixAnalysis} for (a) cases and (b) deaths. These apply hierarchical clustering to the distance $d(s,t)$ between adjacency matrices $\text{Adj}^{(t)}$ at varying times $t$, thereby grouping different \emph{dates} according to the cluster structures at these times. The $y$-axis excludes the first 50 days for cases and 66 days for deaths, as the cluster structure of counts is trivial before these periods, respectively. Each cluster is an unbroken interval of dates. There is a clear break in the cluster structure between 03/01 and 03/02 for cases, and 03/18 and 03/19 for deaths, with a 17-day difference.}
    \label{fig:dendrograms}
\end{figure*}

Finally, we define a distance on the set of \emph{dates} $t=1,\dots,T$. Let the Frobenius norm of an $n \times n$ matrix $A$ be defined as $||A||=\Big(\sum_{i,j=1}^n |a_{ij}|^2\Big) ^{\frac{1}{2}}$. Given $s,t \in [1,\dots, T]$, let $d(s,t)=||\text{Adj}^{(t)} - \text{Adj}^{(s)}||$. Performing hierarchical clustering on these distances $d(s,t)$ produces a dendrogram on the set of dates that we term the \emph{cluster evolution dendrogram}. This groups moments in time according to similarity in the evolving cluster structures. In Appendix \ref{algorithm}, we include an algorithmic presentation of the steps taken in Sections \ref{TimeVaryingClusterAnalysis} and \ref{MatrixAnalysis}. In Appendix \ref{glossary}, we include a list of mathematical objects and their respective definitions used in this paper.

\subsection{Results for time series of cases}
\label{cases results}
In this section, we implement Ckmeans.1d.dp \cite{Wang2011} on daily counts of cases. Experiments using standard K-means on 3-day rolling counts of cases produce similar results included in Appendix \ref{3daysection}. Our analysis supports several aspects of the empirically observed natural history regarding the spread of COVID-19 cases. The smoothed number of clusters $\hat{k}^{(t)}$, depicted in Figure \ref{fig:Offset_cases_deaths}, ranges between $\{2,\dots,17\}$. Until the end of January, there were only two clusters, with China being the only country severely impacted by the virus. However, as the virus has spread around the world, reported counts have changed day by day, with the number of clusters increasing rapidly towards a peak in early March. As depicted in Figure \ref{fig:CasesHeatmap}, Italy was the first country to join the most severely impacted cluster, with the United States (US), Spain, France, Germany, Iran and the United Kingdom (UK) all joining by late March. Subsequently, cluster numbers slowly declined until the end of our analysis window and appear to have stabilized. Indeed, the ranking of worst affected countries has largely stabilized in April, producing more consistent clustering results. 

In Figure \ref{fig:Cases_daily_dendrogram}, we depict the \emph{cluster evolution dendrogram} for the daily cases, defined in Section \ref{MatrixAnalysis}, to study the evolution of the cluster structure. This uses hierarchical clustering to determine similarity between adjacency matrices at different times, which encode the cluster structure on each day. We exclude the first 50 days, in which the cluster structure and associated adjacency matrices are all identical, with only China in its own cluster. The dendrogram identifies two distinct clusters, the larger of which contains two meaningful sub-clusters. All three (sub-)clusters identified are contiguous intervals of dates, 02/19 - 03/01, 03/02 - 03/14, and 03/15 - 04/30. This reveals a marked transition in cluster behaviour on 03/02 for the case counts, with a smaller transition on 03/15.

\subsection{Results for time series of deaths}
\label{deaths results}

In this section, we implement Ckmeans.1d.dp \cite{Wang2011} on daily counts of deaths. The smoothed number of clusters $\hat{k}^{(t)}$, depicted in Figure \ref{fig:Offset_cases_deaths}, ranges between $\{1,\dots,17\}$. The trajectory for number of death clusters follows a similar pattern to that of cases, with a lag of approximately one month. As with the case counts, our analysis highlights the key takeaways in severely impacted countries. Although we have highlighted a one-month offset in the general evolution of COVID-19 cases and deaths, there are dissimilarities regarding the membership of the worst affected cluster. In mid-March, China moved out of the worst cluster into the second death cluster, demonstrating its relative success in responding to the pandemic. On the other hand, the US, Spain, Italy, France and the UK have recently moved into the worst cluster, as depicted in Figure \ref{fig:DeathsHeatmap}. Examining cluster constituencies of cases and deaths over time confirms that China has managed potential COVID-19 deaths relatively effectively, while Italy, Spain, the UK and the US have been ineffective.

In Figure \ref{fig:Deaths_daily_dendrogram}, we depict the \emph{cluster evolution dendrogram}, defined in Section \ref{MatrixAnalysis}, for the daily deaths. We exclude the first 66 days, in which the cluster structure and associated adjacency matrices are all identical. Figures \ref{fig:Cases_daily_dendrogram} and \ref{fig:Deaths_daily_dendrogram} show near-identical hierarchical clustering results for cases and deaths, respectively. Again, two distinct clusters are identified, with two meaningful sub-clusters within the larger cluster. All three (sub-)clusters are again contiguous intervals of dates, 03/06 - 03/18, 03/19 - 03/30 and 03/31 - 04/30. This reveals there is a marked transition in cluster behaviour on 03/19 for the death counts, with a smaller transition on 03/31. These are 17 and 16 days later than the corresponding breaks for the case counts. 


\section{Series offset analysis}
\label{LearningSystemEvolutionOffset}

In this section, we describe further analysis on two related multivariate time series $x_i^{(t)}$ and $y_i^{(t)}$ valued in a common normed space $\mathfrak{X}$. With the application to COVID-19 in mind, we develop a new method that can determine if there an appropriate time offset between the two time series. We perform several analyses for this purpose; in Section \ref{AnomalyAnalysis}, we can subsequently study anomalous individual countries. We adopt our notation from Section \ref{sec: individual}, using subscripts $X$ or $Y$ to refer to mathematical objects pertaining to the cases or deaths counts.

\begin{figure}
\includegraphics[scale=0.48]{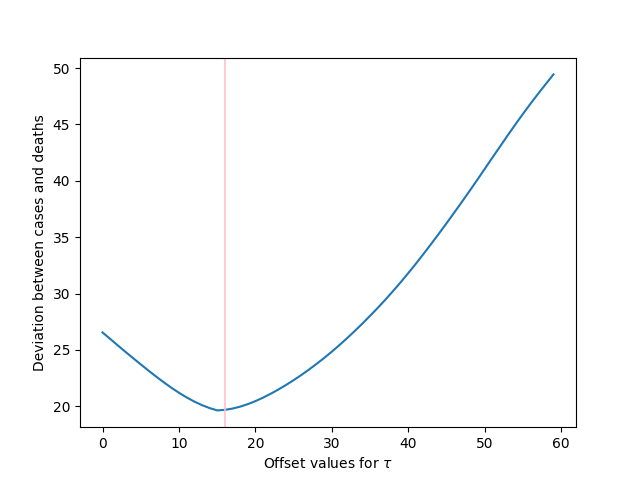}
\caption{\label{fig:epsart} The normalized total offset difference as a function of the offset $\tau$, defined in Eq. (\ref{normalizedmatrixdiff}). The convex nature of this plot indicates that $\tau=16$ is a globally optimal value.}
\end{figure}
\begin{table}
\begin{tabular}{ |p{1.6cm}||p{1.3cm}|p{1.3cm}|p{1.3cm}|p{.7cm}|p{.7cm}|}
 \hline
 \multicolumn{6}{|c|}{Optimal cases vs deaths offset} \\
 \hline
 Start date & Gaussian $m=1$ & Gaussian $m=2$ & Gaussian $m=3$ & Adj & Aff \\
 \hline
31/12/2019 & 16 & 16 & 16 & 20 & 16  \\
 13/1/2020 & 12 & 13 & 14 & 20 & 15 \\
 21/1/2020 & 12 & 13 & 14 & 19 & 15 \\
 31/1/2020 & 12 & 13 & 14 & 19 & 15 \\
\hline
\end{tabular}
\caption{Cluster consistency offset for various adjacency and affinity matrices at different starting dates. These are determined by minimizing the normalized total offset difference in Eq. (\ref{normalizedmatrixdiff}), as well as its analogue for Gaussian and adjacency matrices. The parameter $m$ is defined in Eq. (\ref{affinity equations})}
\label{tab:result_table_days}
\end{table}

First, we have already observed a clear offset in the evolution of $\hat{k}^{(t)}$ for the time series of cases and deaths, and wish to determine it precisely. We define the \emph{series evolution offset} with respect to the changing number of clusters as follows: let $f(t)=\hat{k}_X^{(t)}$ and $g(t)=\hat{k}_Y^{(t)}$ be the smoothed number of clusters for each time series. Given an offset $\delta$, let $f_{\delta}$ be the \emph{translated function} defined by $f_{\delta}(t)=f(t+\delta).$ Let the series evolution offset be the integer $\delta$ that minimizes the $L^1$ distance between functions: 
$$ || f_{\delta} - g||_{L^1} = \int | f_\delta (t) - g(t) | dt$$
For our application, this offset is $\delta=32$, confirming the one-month offset observation in Figure \ref{fig:Offset_cases_deaths}.

\begin{figure*}
    \centering
    \begin{subfigure}[b]{0.32\textwidth}
        \includegraphics[width=\textwidth]{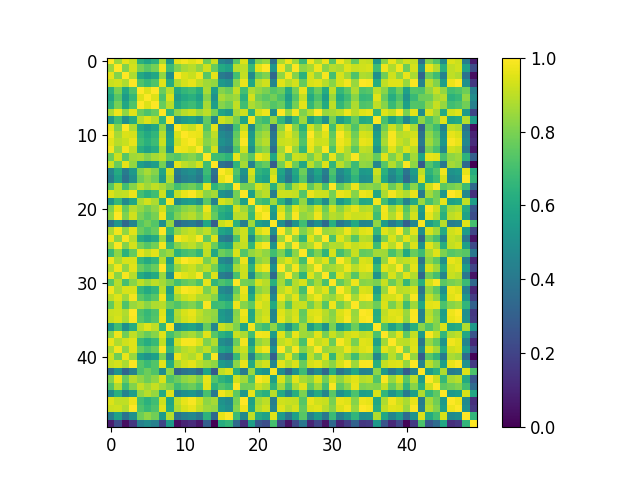}
        \caption{}
        \label{fig:Cases_27_4_20}
    \end{subfigure}
    \begin{subfigure}[b]{0.32\textwidth}
        \includegraphics[width=\textwidth]{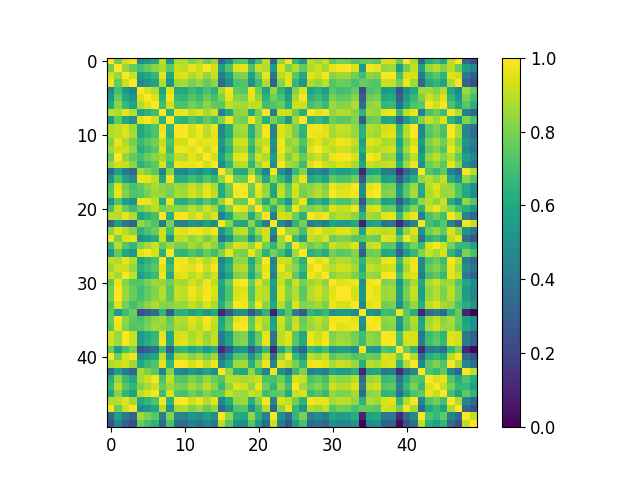}
        \caption{}
        \label{fig:Deaths_27_4_20}
    \end{subfigure}
        \begin{subfigure}[b]{0.32\textwidth}
        \includegraphics[width=\textwidth]{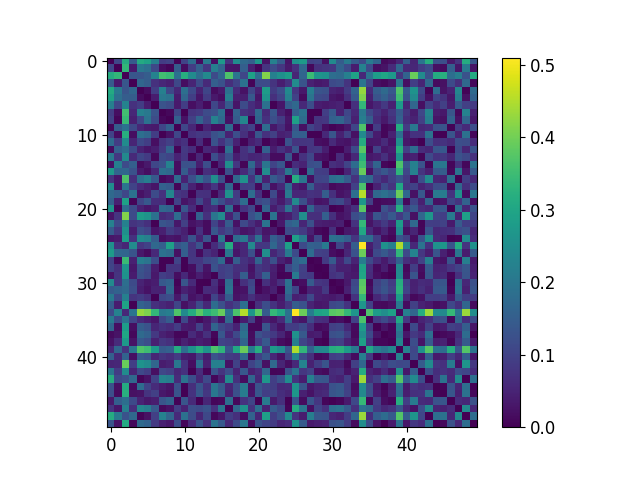}
        \caption{}
        \label{fig:CC_27_4_20}
    \end{subfigure}
    \caption{(a) depicts the affinity matrix for case counts at 4/27/2020, (b) depicts the deaths affinity matrix for 4/11/2020, and (c) depicts the inconsistency matrix with an offset of $\tau=16$ from Table \ref{tab:result_table_days}. Only countries with greater than 5000 cases at 4/30 are included and ordered alphabetically along the axes. The more prominent the respective row and column in the inconsistency matrix, the more anomalous the country. The three most prominent anomalies in (c) are Qatar, Singapore and Bangladesh.}
    \label{CCM}
\end{figure*} 

Next, we determine the offset that minimizes the discrepancy between affinity matrices $\text{Aff}_X$ and $\text{Aff}_Y$ of the two time series. Given an offset $\tau$, let the \emph{normalized total offset difference} between affinity matrices be defined as follows:
\begin{equation}
\frac{1}{T- |\tau|} \sum_{1\leq s,t \leq T, t-s = \tau} ||\text{Aff}^{(s)}_X - \text{Aff}^{(t)}_Y||    
\label{normalizedmatrixdiff}
\end{equation}
We normalize by the number of terms in this sum, which varies with $\tau$, for an appropriate comparison. When $\tau >0$ we can rewrite this as follows: 
$$ \frac{1}{T- \tau} \sum_{t=1}^{T - \tau} ||\text{Aff}^{(t)}_X - \text{Aff}^{(t+\tau)}_Y||  $$
Let  the \emph{cluster consistency offset} be the integer $\tau$ that minimizes the normalized total offset difference. We can also do the same for the offset with respect to the Gaussian affinity or adjacency matrices $G$ and $\text{Adj}$, respectively. All these matrices are normalized, so a comparison of their values is appropriate. We choose the normalization parameter of the Gaussian affinity matrix in Equation (\ref{affinity equations}) for this purpose. We standardize notation such that $\delta$ always refers to the series evolution offset while $\tau$ refers to the cluster consistency offset.

Results are displayed in Table \ref{tab:result_table_days}, with the optimal affinity matrix offset determined in Figure \ref{fig:epsart}. To illustrate the flexibility of the method, we choose different start dates for our offset analysis. The first 30 days carry some triviality in the cluster structure, with few cases observed outside China, so it may be desirable to exclude them from the analysis. Fortunately, the optimal offset differs only slightly with different start dates.

The optimal cluster consistency offset is overwhelmingly around 16. This confirms known medical findings \cite{Lancet} indicating time from diagnosis to death has generally been around 17 days. Moreover, this is consistent with the results of Figure \ref{fig:dendrograms}, where two breaks in the cluster behaviour occurred 17 and 16 days later in the death counts relative to the case counts. This is quite different from the series evolution offset of 32 days. While the cluster consistency offset seeks to align the similarity of case and death counts among individual countries, the series evolution offset seeks to quantify the overall spread of the data as a function of time.


\section{Anomaly analysis}
\label{AnomalyAnalysis}

Having identified a suitable offset between two multivariate time series, one can then investigate the existence of any anomalies. In this case, we use $\tau=16$ as the cluster consistency offset relative to affinity matrices, as depicted in Table \ref{tab:result_table_days}, and then perform a closer analysis of the affinity matrices to identify anomalous countries. Let $\text{Inc}^{(t)}$ be the $n \times n$ \emph{inconsistency matrix} defined entry-wise by $\text{Inc}_{ij}^{(t)} = |\text{Aff}_{X,ij}^{(t)}  - \text{Aff}_{Y,ij}^{(t+\tau)}|$, where the absolute value of each entry is taken. Smaller entries indicate greater consistency between cases and deaths, while greater entries indicate anomalous (inconsistent) countries. Let the \emph{anomaly score} of any individual country be defined as $a^{(t)}_j = \sum_{j=1}^n \text{Inc}^{(t)}_{ij}$. Larger values indicate more anomalous 
countries and the sequence of anomaly scores can reveal the emergence and disappearance of anomalies over time. Let the \emph{lag-adjusted death rate} for each country be defined as follows:
$$ LDR^{(t)}_j = \frac{y_j^{(t)}}{x_j^{(t-\tau)}}, j=1,\dots,n;  t= \tau+1,\dots,T $$
These ratios may be orders of magnitude higher than standard reported death rates, and are no longer bound between 0 and 1. This measure provides insight into the rate of spread, and a country's success in minimizing the number of deaths, conditional on a given number of cases $\tau$  days prior.


In Table \ref{tab:result_table_no}, we depict the results of ordering the 10 most anomalous countries, by anomaly score, from 01/28/2020 - 04/27/2020. In Figure \ref{CCM}, we display the affinity matrices for cases and deaths and the inconsistency matrix for 04/27/2020, with an offset of $\tau=16$ from Table \ref{tab:result_table_days}. We only include countries that had at least 5000 cases as of 04/30/2020. Anomalies may signify either disproportionately high or low number of deaths relative to the number of cases.

\begin{table*}
\tabcolsep 2pt
\begin{tabular}{ |p{2cm}||p{.85cm}|p{.85cm}|p{.85cm}|p{.85cm}|p{.85cm}|p{.85cm}|p{.85cm}|p{.85cm}|p{.85cm}|p{.85cm}|}
 \hline
 \multicolumn{11}{|c|}{10 most anomalous countries: inconsistency matrix analysis} \\
 \hline
 Date & A1 & A2 & A3 & A4 & A5 & A6 & A7 & A8 & A9 & A10\\
 \hline
 \hline	
28/1/2020 & US & UK & IT & IL & IE & IR & ID & IN & DE & FR \\
7/2/2020 & US & DO & IT & IL & IE & IR & ID & IN & DE & FR\\
17/2/2020 & SG & JP & KR & AU & MY & US & DE & FR & AE & CA\\
27/2/2020 & IR & SG & MY & IT & AU & US & DE & UK & AE & CA\\
 8/3/2020 & IT & IR & SG & MY & DE & AE & CA & JP & ES & US\\
 18/3/2020 & ES & SG & IT & IR & AE & UK & NL & FR & US & KR\\
 28/3/2020 & QA & ES & TR & UK & SG & KR & AE & BY & US & IT \\
 7/4/2020 & QA & SG & KR & UK & CN & UA & NO & ZA & AU  & TR\\ 
 17/4/2020 & BD & QA & SG & UK & AU & KR & BE & ZA & AT & FR\\
 27/4/2020 & QA & SG & BD & ME & AU & UK & SW & BE & DE & IL \\
\hline	 
\end{tabular}
\caption{The 10 most anomalous countries in progression from cases to deaths as defined by their anomaly score from Section \ref{AnomalyAnalysis} and a lag of $\tau=16$. AE: United Arab Emirates, AT: Austria, AU: Australia, BD: Bangladesh, BY: Belarus, CA: Canada, CN: China, DE: Germany, DO: Dominican Republic, ES: Spain, FR: France, ID: Indonesia, IE: Ireland, IL: Israel, IN: India, IR: Iran, IT: Italy, JP: Japan, KR: South Korea, MY: Malaysia, NL: Netherlands, NO: Norway, QA: Qatar, SG: Singapore, SW: Sweden, TR: Turkey, UA: Ukraine, UK: United Kingdom, US: United States, ZA: South Africa}
\label{tab:result_table_no}
\end{table*}

This analysis supports several aspects of the empirically observed spread of COVID-19, identifying the most and least successful countries in the progression of cases to deaths. Early in the global spread of COVID-19, Iran and Italy were internationally known as countries that were struggling to contain the number of deaths.\cite{al_jazeera_2020} Table \ref{tab:result_table_no} identifies both as anomalous on 02/27/2020 and 03/08/2020, reflecting their sharp rise in deaths even before other severely impacted countries. On the other hand, Singapore is identified as anomalous during this period due to its relatively small number of deaths. As at 03/07/2020, Singapore had 130 COVID-19 cases and 0 deaths.

A similar trend continued until late March, during which Spain and Italy are identified as the most consistently anomalous countries due to their high death rates. The lag-adjusted death rates for Spain and Italy are  $227\%$ and $73.3\%$, respectively. Indeed, the number of deaths in Spain on 03/28/2020 was more than 2 times greater than the number of cases 16 days earlier. This confirms the severity of the COVID-19 pandemic: Spain and Italy suffered a large number of deaths within a short window. As of late March, Singapore was still identified as anomalous due to the relatively small number of deaths. Towards the end of our analysis window, Qatar and Australia are also identified as anomalous with low death rates, while the UK is identified as anomalous due to a high death rate. The lag-adjusted death rates for Qatar and Australia as of 04/27/2020 are $0.398\%$ and $1.33\%$, respectively. The lag-adjusted death rate for the UK is $34.2\%$. 


\section{Conclusion}
\label{sec: conclusion}


In this paper, we introduce a new method of analyzing a multivariate time series via cluster analysis. Unlike typical applications of time series analysis to epidemiology, it is nonparametric; and unlike existing applications of cluster analysis to time series, we produce a dynamically smoothed number of clusters that changes over time. The analysis of case and death counts over time produces two multivariate time series, which we partition into clusters on each day. While previous studies examine fewer countries over shorter time windows, \cite{Machado2020, Manchein2020} we study 208 countries over four months. Individual countries' cluster membership tracks their severity of counts relative to the rest of the world, while the number of clusters reflects the overall spread of the data. 

The high degree of similarity between the two time series facilitates the identification of anomalous countries in the progression of cases to deaths. We introduce another method herein, using inconsistency matrices and lag-adjusted death rates to highlight the sequential emergence and disappearance of such anomalies over time. These may be used to evaluate a country's effectiveness at handling the pandemic, taking into account an appropriate time offset in mortality due to the disease. Our inconsistency matrices provide a multivariate method with greater generality than the included application. For this reason, they do not identify high or low mortality rates, which are only applicable in a one-dimensional context. The lag-adjusted death rate meets this purpose in our application and any other one-dimensional setting. Lastly, this methodology is flexible: different metrics between data, clustering methods and means of learning offset could all be used to study related multivariate time series and identify changing similarity and anomalies.


Our analysis also provides new insights into the spread of COVID-19 across countries and over time. We show a strong similarity between the evolution of case and death counts, identifying a suitable time offset of 16 days for cluster membership between the two time series. This confirms known medical findings \cite{Lancet} indicating time from diagnosis to death as approximately 17 days. The cluster evolution dendrograms provide further support of a distinct lag between cases and deaths. These dendrograms are highly similar, also up to an offset of 16 days, and demonstrate sharp transition points at 03/02/2020 and 03/19/2020 for cases and deaths, respectively, again with a 17-day difference. These transitions reflect the natural history of the spread of COVID-19 cases and deaths, respectively. On 03/02/2020, numerous countries began to report their first instances of COVID-19 cases, predominantly imported from Iran and Italy. On 03/19/2020, Italy's death toll surpassed that of China. \cite{global_health_2020} Less pronounced transitions exist on 03/15/2020 and 03/31/2020 for cases and deaths, respectively. Again a 16-day offset is observed.

On the other hand, the time offset between the evolution of the number of clusters is 32 days. One explanation for the series evolution offset being longer is that there is an additional delay between cluster membership changes with respect to cases and deaths that can be attributed to stresses on a country's healthcare resources. First, the number of cases may increase significantly, placing a country into a different cluster relative to cases and overwhelming its healthcare resources, thereby leading to a greater number of death counts. That is, the progression from elevation in cases cluster to deaths cluster is not necessarily due to a natural progression from infection to death, but involves mediating factors like stresses on hospital capacity. Perhaps the initial wave of patients can be treated with ventilators, but these may quickly run out, causing more deaths from later instances of cases. Regardless, it is an interesting observation that the offset of 32 days in the number of clusters does not minimize the offset in affinity or adjacency matrix norm differences.


This analysis may assist in identifying the characteristics of the most and least successful government strategies for managing COVID-19. In particular, Singapore, Qatar, Australia and South Korea are four countries whose policies have been most successful in minimizing COVID-19 mortality. Each of these countries provided a substantial amount of easily accessible testing in the early stages of COVID-19 development.\cite{guardian_2020} Singapore and Australia also closed their borders to travel before a critical mass in total case counts was established, and were early to implement strict lockdown procedures. \cite{bbc_2020} 

By contrast, Italy, Spain and the UK are three countries whose policies managed the progression from COVID-19 cases to deaths least effectively. Many argue that lockdown procedures in Italy and Spain, although severe once in place, were implemented too late. \cite{nyt2020} Similarly, the UK initially elected not to shut down large gatherings or introduce social distancing measures in an attempt to build herd immunity among the community. Ultimately, however, the UK did implement strict lockdown policies as mortality rates rose. \cite{fp_2020}

These findings suggest that the timeliness of various lockdown procedures is perhaps more important than their severity. Countries with easy access to early testing also appear to manage the progression from cases to deaths more effectively. Conversely, countries that struggled to minimize their COVID-19 mortality rate also exhibit some general similarities. First, these countries were slow to implement measures that would restrict people's movements. Second,  many of these countries carried an early high case burden, suggesting that mediating factors such as undue stress from finite healthcare resources may contribute to the mortality rate.

Overall, this paper introduces a new method for analyzing multivariate time series individually and in conjunction, thereby providing new insights into the caseload and mortality rate affecting different countries. As the pandemic evolves, it is the objective of emerging research to facilitate timely and appropriate means of producing effective government strategies for minimizing the extensive human, social and cultural costs of COVID-19.

\section*{Data availability}
The data that support the findings of this study are openly available at Ref. \onlinecite{worldindata2020}.

\begin{acknowledgments}
The authors thank Kerry Chen and Alex Judge for helpful comments and edits.
\end{acknowledgments}

\appendix

\section{Existing cluster theory}
\label{existingclustertheory}
In this section, we provide an overview of the three clustering algorithms used in the body of the paper: hierarchical clustering, K-means, and its  optimal one-dimensional variant Ckmeans.1d.dp. In our most general setup, $x_1,\dots,x_n$ are elements of a normed space $\mathfrak{X}$. 

\emph{Hierarchical clustering}\cite{Ward1963, Szekely2005} is an iterative clustering technique that does not specify discrete groupings of elements. Rather, it seeks to build a hierarchy of similarity between elements. Hierarchical clustering is either agglomerative, where each element $x_i$ begins in its own cluster and branches between them are successively built, or divisive, where all elements begin in one cluster and are successively split. The results of hierarchical clustering are commonly displayed in \emph{dendrograms}. Hierarchical clustering does not require the choice of a number of clusters $k$. In this paper, hierarchical clustering is exclusively used to produce the dendrograms of Figure \ref{fig:dendrograms}. There, we implement agglomerative clustering.

\emph{K-means clustering} seeks to minimize an appropriate sum of square distances. With $k$ chosen \textit{a priori}, we investigate all possible partitions (disjoint unions)  $C_1 \cup C_2 \cup \dots \cup C_k $ of $\{ x_1,\dots, x_n \}$. Let $z_j$ be the \emph{centroid} (average) of the subset $C_j$. One seeks to minimize the sum of square distances within each cluster to its centroid:
$$ \sum_{j=1}^k \sum_{x \in C_j} ||x - z_j||^2 $$
For a normed space with dimension at least 2, it is NP-hard to find the global optimum of this problem. The K-means algorithm  \cite{Lloyd1982} is an iterative algorithm that converges quickly and suitably to a locally optimal solution. It is usually sufficient for applications. In this paper, multivariate K-means is exclusively used in Figure \ref{fig:3dayheatmaps}.

On the other hand, the K-means optimisation problem is efficiently solvable in the one-dimensional case - when $x_i$ are real numbers, they are equipped with an ordering, which considerably simplifies the problem. To cluster $n$ elements of $\mathfrak{X}=\mathbb{R}$ into $k$ clusters requires one to order the elements and then determine $k-1$ breaks in the ordering. This is far less computationally intensive than the higher-dimensional analogue. Ckmeans.1d.dp \cite{Wang2011} is a dynamic programming algorithm that guarantees optimal clustering in one dimension, choosing $k$ \textit{a priori}. 

How to best choose the number of clusters $k$ for the K-means algorithm is a difficult problem. Different methods for estimating $k$ may produce considerably differing results. In this paper, we draw upon six methods to determine the appropriate number of clusters before implementing K-means, in both the one and higher-dimensional cases. These methods are well-known: Ptbiserial index \cite{Milligan1980}, silhouette score \cite{Rousseeuw1987}, KL index \cite{krzanowski1988}, C index \cite{Hubert1976}, McClain-Rao index \cite{Mcclain1975} and Dunn index \cite{Dunn1974}. We have chosen these methods based upon consultation with the literature and our own experiments. However, our methodology is flexible, and any combination of existing methods may be used. For one-dimensional data, it is often regarded as unsuitable to use multivariate clustering methods, as optimal alternatives exist. Since we study one-dimensional data in this paper, it is necessary to use these methods to choose the number $k$ before implementation of Ckmeans.1d.dp.

In the body of the paper, we choose the smoothed number of clusters $\hat{k}^{(t)}$, depicted in Figure \ref{fig:movingk}, by applying exponential smoothing to the average of the six choices of cluster number listed above. We then apply Ckmeans.1d.dp to divide daily counts of data into $\hat{k}^{(t)}$ clusters. This determines our results in Figure \ref{fig:heatmaps}. In Figure \ref{fig:3dayheatmaps}, we display analogous results for 3-day rolling counts, clustering the corresponding elements of $\mathbb{R}^3$ using standard K-means. The results are highly similar.

\section{3-day rolling counts}
\label{3daysection}

\begin{figure}[H]
    \centering
    \begin{subfigure}[b]{0.45\textwidth}
        \includegraphics[width=\textwidth]{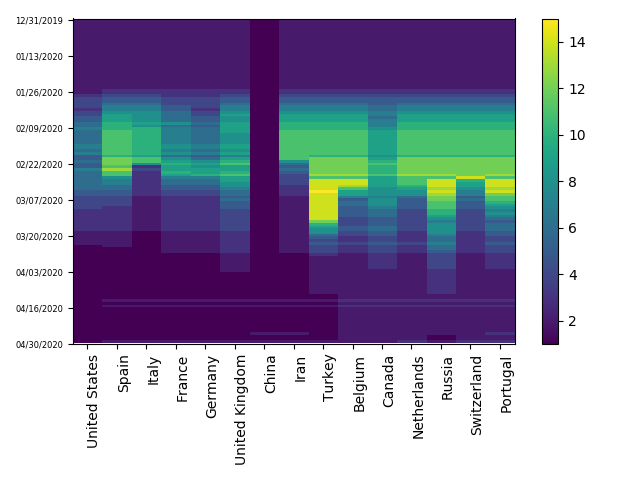}
        \caption{}
    \label{fig:3dCasesHeatmap}
    \end{subfigure}
    \begin{subfigure}[b]{0.45\textwidth}
        \includegraphics[width=\textwidth]{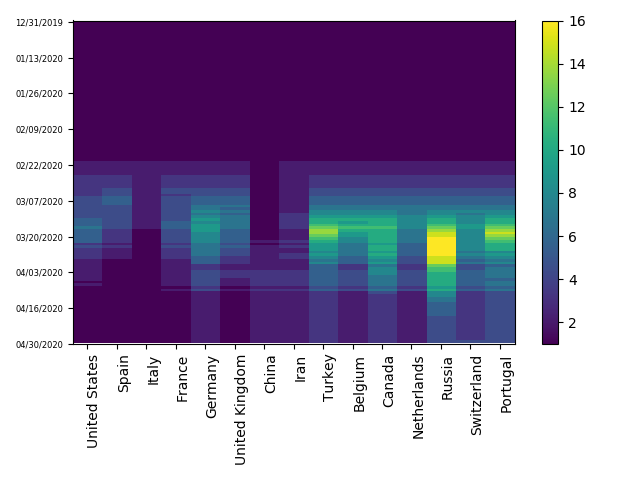}
        \caption{}
    \label{fig:3dDeathsHeatmap}
    \end{subfigure}
    \caption{Heat maps track the changing cluster membership of the fifteen most severely impacted countries with respect to their 3-day rolling counts of (a) cases and (b) deaths, respectively. Cluster membership, determined by K-means, depicts COVID-19 severity relative to the rest of the world. There is strong similarity relative to Figure \ref{fig:heatmaps}. }
    \label{fig:3dayheatmaps}
\end{figure}

In this section, we briefly show the applicability of our method to higher-dimensional data. We present two multivariate time series of the cumulative 3-day rolling counts of cases and deaths on a country by country basis. We order the countries by alphabetical order and denote these 3-day rolling counts by $\mathbf{\tilde{x}}_i^{(t)}, \mathbf{\tilde{y}}_i^{(t)} \in \mathbb{R}^3, i=1,\dots,208$. We proceed exactly as in Section \ref{sec: individual}, applying standard K-means instead of Ckmeans.1d.dp. In Figure \ref{fig:3dayheatmaps}, we depict the same countries' changing cluster membership as were depicted in Figure \ref{fig:heatmaps}. The similarity shows the robustness and generality of our method.

\section{Algorithmic description of methodology}
\label{algorithm}

In this section, we provide an algorithmic presentation of the computational steps taken for the analysis of an individual multivariate time series, described in Sections \ref{TimeVaryingClusterAnalysis} and \ref{MatrixAnalysis}.

\begin{algorithm}[H]
    \renewcommand{\thealgorithm}{}
	\caption{Cluster-based evolution analysis}
	\begin{algorithmic} 
	\State Given: a multivariate time series $x^{(t)}_{i} \in \mathbb{R}_{\geq 0}$
	\State Data preprocessing:
	\If {$x^{(t)}_{i} = 0$ or NaN}
	\State $x^{(t)}_{i} = 1$
	\EndIf
	\State Data transformation:
	\State $x^{(t)}_{i} = \log x^{(t)}_{i}$
	\For{$t=1 \,\text{to}\, T$}
	\State Compute $k^{(t)}_1,...,k^{(t)}_6$
	\State $k^{t}_1 = \text{Ptbiserial}\bigg((x_{i:1,N}^{(t)})_{1 \leq i \leq N}\bigg)$
	\State $k^{(t)}_2 = \text{Silhouette score}\bigg((x_{i:1,N}^{(t)})_{1 \leq i \leq N}\bigg)$
	\State $k^{(t)}_3 = \text{KL index}\bigg((x_{i:1,N}^{(t)})_{1 \leq i \leq N}\bigg)$
	\State $k^{(t)}_4 = \text{C index}\bigg((x_{i:1,N}^{(t)})_{1 \leq i \leq N}\bigg)$
	\State $k^{(t)}_5 = \text{McClain-Rao index}\bigg((x_{i:1,N}^{(t)})_{1 \leq i \leq N}\bigg)$
	\State $k^{(t)}_6 = \text{Dunn index}\bigg((x_{i:1,N}^{(t)})_{1 \leq i \leq N}\bigg)$
	\State $k^{(t)}_{av} = \frac{1}{6} \sum k^{(t)}_{i}$
	\EndFor
	\State \textbf{End for}
	\State $\hat{k}^{(t)} = \text{simple exponential smoothing}(k_{av}^{(t)})$
	\For{$t=1 \,\text{to}\, T$}
	\State $\text{Ckmeans.1d.dp sort} (x^{(t)}_{i}) \text{ into } \hat{k}^{(t)}$ clusters
	\State Record and sort cluster labels.
	\State Let $\text{Adj}^{(t)}_{ij}$ be adjacency matrix.
	\If{$x^{(t)}_{i}$ and $x^{(t)}_{j}$ are in same cluster} \State $\text{Adj}^{(t)}_{ij} = 1$
	\Else 
	\State $\text{Adj}^{(t)}_{ij} = 0$
	\EndIf
	\EndFor
	\State \textbf{End for}
	\State Compute affinity matrix, $\text{Aff}^{(t)}_{ij} = 1 - \frac{D^{(t)}_{ij}}{\max D^{(t)}}$ 
	\State Compute Gaussian matrix $G^{(t)}_{ij} =\exp \bigg(\frac{-m^2\big(D^{(t)}_{ij}\big)^2}{2(\max D^{(t)})^2} \bigg)$ 
	\State Compute  $d(s,t)=||\text{Adj}^{(t)} - \text{Adj}^{(s)}||$.
	\State \textbf{do} hierarchical clustering on $d(s,t), 1 \leq s,t \leq T$.
	\end{algorithmic}
	
\end{algorithm}

\section{Glossary of mathematical objects}
\label{mathematics_objects_glossary}

In this brief section, we include a glossary of mathematical objects presented in the paper and their respective definitions in Table \ref{tab:MathematicalObjects}.

\label{glossary}
\begin{table}[H]
\begin{tabular}{ |p{1.4cm}||p{6.8cm}|}
 \hline
 \multicolumn{2}{|c|}{\textbf{Mathematical objects glossary}} \\
 \hline
 Object & Description \\
 \hline
 $D^{(t)}$ & Distance matrix between log counts \\
$\text{Aff}^{(t)}$ & Standard affinity matrix \\
$G^{(t)}$ & Gaussian affinity matrix \\
 $k_{av}^{(t)}$ & Unsmoothed number of clusters obtained as average of six methods \\
 $\hat{k}^{(t)}$ & Smoothed number of clusters \\
 $\text{Adj}^{(t)}$ & Adjacency matrix coding cluster outputs for $\hat{k}^{(t)}$ clusters  \\
$d(s,t)$ & Frobenius distance between adjacency matrix of various dates \\
$\delta$ & Series evolution offset with respect to number of clusters \\
$\tau$ & Cluster consistency offset with respect to cluster membership \\
$\text{Inc}^{(t)}$ & Lag-adjusted inconsistency matrix \\
$a_j^{(t)}$ & Anomaly score of country $j$ \\
$LDR^{(t)}_j$ & Lag-adjusted death rate of country $j$  \\
$||f_\delta - g||_{L^1}$ & $L^1$ norm between functions \\
\hline
\end{tabular}
\caption{Mathematical objects and definitions}
\label{tab:MathematicalObjects}
\end{table}


\bibliography{aipsamp}

\begin{thebibliography}{35}%
\makeatletter
\providecommand \@ifxundefined [1]{%
 \@ifx{#1\undefined}
}%
\providecommand \@ifnum [1]{%
 \ifnum #1\expandafter \@firstoftwo
 \else \expandafter \@secondoftwo
 \fi
}%
\providecommand \@ifx [1]{%
 \ifx #1\expandafter \@firstoftwo
 \else \expandafter \@secondoftwo
 \fi
}%
\providecommand \natexlab [1]{#1}%
\providecommand \enquote  [1]{``#1''}%
\providecommand \bibnamefont  [1]{#1}%
\providecommand \bibfnamefont [1]{#1}%
\providecommand \citenamefont [1]{#1}%
\providecommand \href@noop [0]{\@secondoftwo}%
\providecommand \href [0]{\begingroup \@sanitize@url \@href}%
\providecommand \@href[1]{\@@startlink{#1}\@@href}%
\providecommand \@@href[1]{\endgroup#1\@@endlink}%
\providecommand \@sanitize@url [0]{\catcode `\\12\catcode `\$12\catcode
  `\&12\catcode `\#12\catcode `\^12\catcode `\_12\catcode `\%12\relax}%
\providecommand \@@startlink[1]{}%
\providecommand \@@endlink[0]{}%
\providecommand \url  [0]{\begingroup\@sanitize@url \@url }%
\providecommand \@url [1]{\endgroup\@href {#1}{\urlprefix }}%
\providecommand \urlprefix  [0]{URL }%
\providecommand \Eprint [0]{\href }%
\providecommand \doibase [0]{http://dx.doi.org/}%
\providecommand \selectlanguage [0]{\@gobble}%
\providecommand \bibinfo  [0]{\@secondoftwo}%
\providecommand \bibfield  [0]{\@secondoftwo}%
\providecommand \translation [1]{[#1]}%
\providecommand \BibitemOpen [0]{}%
\providecommand \bibitemStop [0]{}%
\providecommand \bibitemNoStop [0]{.\EOS\space}%
\providecommand \EOS [0]{\spacefactor3000\relax}%
\providecommand \BibitemShut  [1]{\csname bibitem#1\endcsname}%
\let\auto@bib@innerbib\@empty
\bibitem [{\citenamefont {Hethcote}(2000)}]{Hethcote2000}%
  \BibitemOpen
  \bibfield  {author} {\bibinfo {author} {\bibfnamefont {H.~W.}\ \bibnamefont
  {Hethcote}},\ }\bibfield  {title} {\enquote {\bibinfo {title} {The
  mathematics of infectious diseases},}\ }\href {\doibase
  10.1137/s0036144500371907} {\bibfield  {journal} {\bibinfo  {journal} {{SIAM}
  Review}\ }\textbf {\bibinfo {volume} {42}},\ \bibinfo {pages} {599--653}
  (\bibinfo {year} {2000})}\BibitemShut {NoStop}%
\bibitem [{\citenamefont {Chowell}\ \emph {et~al.}(2016)\citenamefont
  {Chowell}, \citenamefont {Sattenspiel}, \citenamefont {Bansal},\ and\
  \citenamefont {Viboud}}]{Chowell2016}%
  \BibitemOpen
  \bibfield  {author} {\bibinfo {author} {\bibfnamefont {G.}~\bibnamefont
  {Chowell}}, \bibinfo {author} {\bibfnamefont {L.}~\bibnamefont
  {Sattenspiel}}, \bibinfo {author} {\bibfnamefont {S.}~\bibnamefont {Bansal}},
  \ and\ \bibinfo {author} {\bibfnamefont {C.}~\bibnamefont {Viboud}},\
  }\bibfield  {title} {\enquote {\bibinfo {title} {Mathematical models to
  characterize early epidemic growth: A review},}\ }\href {\doibase
  10.1016/j.plrev.2016.07.005} {\bibfield  {journal} {\bibinfo  {journal}
  {Physics of Life Reviews}\ }\textbf {\bibinfo {volume} {18}},\ \bibinfo
  {pages} {66--97} (\bibinfo {year} {2016})}\BibitemShut {NoStop}%
\bibitem [{\citenamefont {Vazquez}(2006)}]{Vazquez2006}%
  \BibitemOpen
  \bibfield  {author} {\bibinfo {author} {\bibfnamefont {A.}~\bibnamefont
  {Vazquez}},\ }\bibfield  {title} {\enquote {\bibinfo {title} {Polynomial
  growth in branching processes with diverging reproductive number},}\ }\href
  {\doibase 10.1103/physrevlett.96.038702} {\bibfield  {journal} {\bibinfo
  {journal} {Physical Review Letters}\ }\textbf {\bibinfo {volume} {96}}
  (\bibinfo {year} {2006}),\ 10.1103/physrevlett.96.038702}\BibitemShut
  {NoStop}%
\bibitem [{\citenamefont {Moeckel}\ and\ \citenamefont
  {Murray}(1997)}]{Moeckel1997}%
  \BibitemOpen
  \bibfield  {author} {\bibinfo {author} {\bibfnamefont {R.}~\bibnamefont
  {Moeckel}}\ and\ \bibinfo {author} {\bibfnamefont {B.}~\bibnamefont
  {Murray}},\ }\bibfield  {title} {\enquote {\bibinfo {title} {Measuring the
  distance between time series},}\ }\href@noop {} {\bibfield  {journal}
  {\bibinfo  {journal} {Physica D}\ } (\bibinfo {year} {1997})}\BibitemShut
  {NoStop}%
\bibitem [{\citenamefont {Sz{\'{e}}kely}, \citenamefont {Rizzo},\ and\
  \citenamefont {Bakirov}(2007)}]{Szkely2007}%
  \BibitemOpen
  \bibfield  {author} {\bibinfo {author} {\bibfnamefont {G.~J.}\ \bibnamefont
  {Sz{\'{e}}kely}}, \bibinfo {author} {\bibfnamefont {M.~L.}\ \bibnamefont
  {Rizzo}}, \ and\ \bibinfo {author} {\bibfnamefont {N.~K.}\ \bibnamefont
  {Bakirov}},\ }\bibfield  {title} {\enquote {\bibinfo {title} {Measuring and
  testing dependence by correlation of distances},}\ }\href {\doibase
  10.1214/009053607000000505} {\bibfield  {journal} {\bibinfo  {journal} {The
  Annals of Statistics}\ }\textbf {\bibinfo {volume} {35}},\ \bibinfo {pages}
  {2769--2794} (\bibinfo {year} {2007})}\BibitemShut {NoStop}%
\bibitem [{\citenamefont {Mendes}\ and\ \citenamefont
  {Beims}(2018)}]{Mendes2018}%
  \BibitemOpen
  \bibfield  {author} {\bibinfo {author} {\bibfnamefont {C.~F.}\ \bibnamefont
  {Mendes}}\ and\ \bibinfo {author} {\bibfnamefont {M.~W.}\ \bibnamefont
  {Beims}},\ }\bibfield  {title} {\enquote {\bibinfo {title} {Distance
  correlation detecting {L}yapunov instabilities, noise-induced escape times
  and mixing},}\ }\href {\doibase 10.1016/j.physa.2018.08.028} {\bibfield
  {journal} {\bibinfo  {journal} {Physica A: Statistical Mechanics and its
  Applications}\ }\textbf {\bibinfo {volume} {512}},\ \bibinfo {pages}
  {721--730} (\bibinfo {year} {2018})}\BibitemShut {NoStop}%
\bibitem [{\citenamefont {Mendes}, \citenamefont {da~Silva},\ and\
  \citenamefont {Beims}(2019)}]{Mendes2019}%
  \BibitemOpen
  \bibfield  {author} {\bibinfo {author} {\bibfnamefont {C.~F.~O.}\
  \bibnamefont {Mendes}}, \bibinfo {author} {\bibfnamefont {R.~M.}\
  \bibnamefont {da~Silva}}, \ and\ \bibinfo {author} {\bibfnamefont {M.~W.}\
  \bibnamefont {Beims}},\ }\bibfield  {title} {\enquote {\bibinfo {title}
  {Decay of the distance autocorrelation and {L}yapunov exponents},}\ }\href
  {\doibase 10.1103/physreve.99.062206} {\bibfield  {journal} {\bibinfo
  {journal} {Physical Review E}\ }\textbf {\bibinfo {volume} {99}} (\bibinfo
  {year} {2019}),\ 10.1103/physreve.99.062206}\BibitemShut {NoStop}%
\bibitem [{\citenamefont {Shang}\ \emph {et~al.}(2020)\citenamefont {Shang},
  \citenamefont {Yang}, \citenamefont {Moore}, \citenamefont {Ji},\ and\
  \citenamefont {Small}}]{Shang2020}%
  \BibitemOpen
  \bibfield  {author} {\bibinfo {author} {\bibfnamefont {K.}~\bibnamefont
  {Shang}}, \bibinfo {author} {\bibfnamefont {B.}~\bibnamefont {Yang}},
  \bibinfo {author} {\bibfnamefont {J.~M.}\ \bibnamefont {Moore}}, \bibinfo
  {author} {\bibfnamefont {Q.}~\bibnamefont {Ji}}, \ and\ \bibinfo {author}
  {\bibfnamefont {M.}~\bibnamefont {Small}},\ }\bibfield  {title} {\enquote
  {\bibinfo {title} {Growing networks with communities: A distributive link
  model},}\ }\href {\doibase 10.1063/5.0007422} {\bibfield  {journal} {\bibinfo
   {journal} {Chaos: An Interdisciplinary Journal of Nonlinear Science}\
  }\textbf {\bibinfo {volume} {30}},\ \bibinfo {pages} {041101} (\bibinfo
  {year} {2020})}\BibitemShut {NoStop}%
\bibitem [{\citenamefont {Manchein}\ \emph {et~al.}(2020)\citenamefont
  {Manchein}, \citenamefont {Brugnago}, \citenamefont {da~Silva}, \citenamefont
  {Mendes},\ and\ \citenamefont {Beims}}]{Manchein2020}%
  \BibitemOpen
  \bibfield  {author} {\bibinfo {author} {\bibfnamefont {C.}~\bibnamefont
  {Manchein}}, \bibinfo {author} {\bibfnamefont {E.~L.}\ \bibnamefont
  {Brugnago}}, \bibinfo {author} {\bibfnamefont {R.~M.}\ \bibnamefont
  {da~Silva}}, \bibinfo {author} {\bibfnamefont {C.~F.~O.}\ \bibnamefont
  {Mendes}}, \ and\ \bibinfo {author} {\bibfnamefont {M.~W.}\ \bibnamefont
  {Beims}},\ }\bibfield  {title} {\enquote {\bibinfo {title} {Strong
  correlations between power-law growth of {COVID}-19 in four continents and
  the inefficiency of soft quarantine strategies},}\ }\href {\doibase
  10.1063/5.0009454} {\bibfield  {journal} {\bibinfo  {journal} {Chaos: An
  Interdisciplinary Journal of Nonlinear Science}\ }\textbf {\bibinfo {volume}
  {30}},\ \bibinfo {pages} {041102} (\bibinfo {year} {2020})}\BibitemShut
  {NoStop}%
\bibitem [{\citenamefont {Machado}\ and\ \citenamefont
  {Lopes}(2020)}]{Machado2020}%
  \BibitemOpen
  \bibfield  {author} {\bibinfo {author} {\bibfnamefont {J.~A.~T.}\
  \bibnamefont {Machado}}\ and\ \bibinfo {author} {\bibfnamefont {A.~M.}\
  \bibnamefont {Lopes}},\ }\bibfield  {title} {\enquote {\bibinfo {title} {Rare
  and extreme events: the case of {COVID}-19 pandemic},}\ }\href {\doibase
  10.1007/s11071-020-05680-w} {\bibfield  {journal} {\bibinfo  {journal}
  {Nonlinear Dynamics}\ } (\bibinfo {year} {2020}),\
  10.1007/s11071-020-05680-w}\BibitemShut {NoStop}%
\bibitem [{\citenamefont {Cassetti}\ \emph {et~al.}(2008)\citenamefont
  {Cassetti}, \citenamefont {Rosa}, \citenamefont {Rossi}, \citenamefont
  {D'Al{\`{o}}},\ and\ \citenamefont {Stracci}}]{Cassetti2008}%
  \BibitemOpen
  \bibfield  {author} {\bibinfo {author} {\bibfnamefont {T.}~\bibnamefont
  {Cassetti}}, \bibinfo {author} {\bibfnamefont {F.~L.}\ \bibnamefont {Rosa}},
  \bibinfo {author} {\bibfnamefont {L.}~\bibnamefont {Rossi}}, \bibinfo
  {author} {\bibfnamefont {D.}~\bibnamefont {D'Al{\`{o}}}}, \ and\ \bibinfo
  {author} {\bibfnamefont {F.}~\bibnamefont {Stracci}},\ }\bibfield  {title}
  {\enquote {\bibinfo {title} {Cancer incidence in men: a cluster analysis of
  spatial patterns},}\ }\href {\doibase 10.1186/1471-2407-8-344} {\bibfield
  {journal} {\bibinfo  {journal} {{BMC} Cancer}\ }\textbf {\bibinfo {volume}
  {8}} (\bibinfo {year} {2008}),\ 10.1186/1471-2407-8-344}\BibitemShut
  {NoStop}%
\bibitem [{\citenamefont {Alashwal}\ \emph {et~al.}(2019)\citenamefont
  {Alashwal}, \citenamefont {Halaby}, \citenamefont {Crouse}, \citenamefont
  {Abdalla},\ and\ \citenamefont {Moustafa}}]{Alashwal2019}%
  \BibitemOpen
  \bibfield  {author} {\bibinfo {author} {\bibfnamefont {H.}~\bibnamefont
  {Alashwal}}, \bibinfo {author} {\bibfnamefont {M.~E.}\ \bibnamefont
  {Halaby}}, \bibinfo {author} {\bibfnamefont {J.~J.}\ \bibnamefont {Crouse}},
  \bibinfo {author} {\bibfnamefont {A.}~\bibnamefont {Abdalla}}, \ and\
  \bibinfo {author} {\bibfnamefont {A.~A.}\ \bibnamefont {Moustafa}},\
  }\bibfield  {title} {\enquote {\bibinfo {title} {The application of
  unsupervised clustering methods to {A}lzheimer's disease},}\ }\href {\doibase
  10.3389/fncom.2019.00031} {\bibfield  {journal} {\bibinfo  {journal}
  {Frontiers in Computational Neuroscience}\ }\textbf {\bibinfo {volume} {13}}
  (\bibinfo {year} {2019}),\ 10.3389/fncom.2019.00031}\BibitemShut {NoStop}%
\bibitem [{\citenamefont {Xiao}\ \emph {et~al.}(2016)\citenamefont {Xiao},
  \citenamefont {van Hoek}, \citenamefont {Kenward}, \citenamefont {Melegaro},\
  and\ \citenamefont {Jit}}]{Xiao2016}%
  \BibitemOpen
  \bibfield  {author} {\bibinfo {author} {\bibfnamefont {X.}~\bibnamefont
  {Xiao}}, \bibinfo {author} {\bibfnamefont {A.~J.}\ \bibnamefont {van Hoek}},
  \bibinfo {author} {\bibfnamefont {M.~G.}\ \bibnamefont {Kenward}}, \bibinfo
  {author} {\bibfnamefont {A.}~\bibnamefont {Melegaro}}, \ and\ \bibinfo
  {author} {\bibfnamefont {M.}~\bibnamefont {Jit}},\ }\bibfield  {title}
  {\enquote {\bibinfo {title} {Clustering of contacts relevant to the spread of
  infectious disease},}\ }\href {\doibase 10.1016/j.epidem.2016.08.001}
  {\bibfield  {journal} {\bibinfo  {journal} {Epidemics}\ }\textbf {\bibinfo
  {volume} {17}},\ \bibinfo {pages} {1--9} (\bibinfo {year}
  {2016})}\BibitemShut {NoStop}%
\bibitem [{\citenamefont {McCloskey}\ and\ \citenamefont
  {Poon}(2017)}]{McCloskey2017}%
  \BibitemOpen
  \bibfield  {author} {\bibinfo {author} {\bibfnamefont {R.~M.}\ \bibnamefont
  {McCloskey}}\ and\ \bibinfo {author} {\bibfnamefont {A.~F.~Y.}\ \bibnamefont
  {Poon}},\ }\bibfield  {title} {\enquote {\bibinfo {title} {A model-based
  clustering method to detect infectious disease transmission outbreaks from
  sequence variation},}\ }\href {\doibase 10.1371/journal.pcbi.1005868}
  {\bibfield  {journal} {\bibinfo  {journal} {{PLOS} Computational Biology}\
  }\textbf {\bibinfo {volume} {13}},\ \bibinfo {pages} {e1005868} (\bibinfo
  {year} {2017})}\BibitemShut {NoStop}%
\bibitem [{\citenamefont {Muradi}, \citenamefont {Bustamam},\ and\
  \citenamefont {Lestari}(2015)}]{Muradi2015}%
  \BibitemOpen
  \bibfield  {author} {\bibinfo {author} {\bibfnamefont {H.}~\bibnamefont
  {Muradi}}, \bibinfo {author} {\bibfnamefont {A.}~\bibnamefont {Bustamam}}, \
  and\ \bibinfo {author} {\bibfnamefont {D.}~\bibnamefont {Lestari}},\
  }\bibfield  {title} {\enquote {\bibinfo {title} {Application of hierarchical
  clustering ordered partitioning and collapsing hybrid in {E}bola virus
  phylogenetic analysis},}\ }in\ \href {\doibase 10.1109/icacsis.2015.7415183}
  {\emph {\bibinfo {booktitle} {2015 International Conference on Advanced
  Computer Science and Information Systems ({ICACSIS})}}}\ (\bibinfo
  {publisher} {{IEEE}},\ \bibinfo {year} {2015})\BibitemShut {NoStop}%
\bibitem [{\citenamefont {Rizzi}\ \emph {et~al.}(2010)\citenamefont {Rizzi},
  \citenamefont {Mahata}, \citenamefont {Mathieson},\ and\ \citenamefont
  {Moscato}}]{Rizzi2010}%
  \BibitemOpen
  \bibfield  {author} {\bibinfo {author} {\bibfnamefont {R.}~\bibnamefont
  {Rizzi}}, \bibinfo {author} {\bibfnamefont {P.}~\bibnamefont {Mahata}},
  \bibinfo {author} {\bibfnamefont {L.}~\bibnamefont {Mathieson}}, \ and\
  \bibinfo {author} {\bibfnamefont {P.}~\bibnamefont {Moscato}},\ }\bibfield
  {title} {\enquote {\bibinfo {title} {Hierarchical clustering using the
  arithmetic-harmonic cut: Complexity and experiments},}\ }\href {\doibase
  10.1371/journal.pone.0014067} {\bibfield  {journal} {\bibinfo  {journal}
  {{PLoS} {ONE}}\ }\textbf {\bibinfo {volume} {5}},\ \bibinfo {pages} {e14067}
  (\bibinfo {year} {2010})}\BibitemShut {NoStop}%
\bibitem [{\citenamefont {Lloyd}(1982)}]{Lloyd1982}%
  \BibitemOpen
  \bibfield  {author} {\bibinfo {author} {\bibfnamefont {S.}~\bibnamefont
  {Lloyd}},\ }\bibfield  {title} {\enquote {\bibinfo {title} {Least squares
  quantization in {PCM}},}\ }\href {\doibase 10.1109/tit.1982.1056489}
  {\bibfield  {journal} {\bibinfo  {journal} {{IEEE} Transactions on
  Information Theory}\ }\textbf {\bibinfo {volume} {28}},\ \bibinfo {pages}
  {129--137} (\bibinfo {year} {1982})}\BibitemShut {NoStop}%
\bibitem [{\citenamefont {von Luxburg}(2007)}]{Luxburg2007}%
  \BibitemOpen
  \bibfield  {author} {\bibinfo {author} {\bibfnamefont {U.}~\bibnamefont {von
  Luxburg}},\ }\bibfield  {title} {\enquote {\bibinfo {title} {A tutorial on
  spectral clustering},}\ }\href {\doibase 10.1007/s11222-007-9033-z}
  {\bibfield  {journal} {\bibinfo  {journal} {Statistics and Computing}\
  }\textbf {\bibinfo {volume} {17}},\ \bibinfo {pages} {395--416} (\bibinfo
  {year} {2007})}\BibitemShut {NoStop}%
\bibitem [{\citenamefont {Ward}(1963)}]{Ward1963}%
  \BibitemOpen
  \bibfield  {author} {\bibinfo {author} {\bibfnamefont {J.~H.}\ \bibnamefont
  {Ward}},\ }\bibfield  {title} {\enquote {\bibinfo {title} {Hierarchical
  grouping to optimize an objective function},}\ }\href {\doibase
  10.1080/01621459.1963.10500845} {\bibfield  {journal} {\bibinfo  {journal}
  {Journal of the {A}merican Statistical Association}\ }\textbf {\bibinfo
  {volume} {58}},\ \bibinfo {pages} {236--244} (\bibinfo {year}
  {1963})}\BibitemShut {NoStop}%
\bibitem [{\citenamefont {Szekely}\ and\ \citenamefont
  {Rizzo}(2005)}]{Szekely2005}%
  \BibitemOpen
  \bibfield  {author} {\bibinfo {author} {\bibfnamefont {G.~J.}\ \bibnamefont
  {Szekely}}\ and\ \bibinfo {author} {\bibfnamefont {M.~L.}\ \bibnamefont
  {Rizzo}},\ }\bibfield  {title} {\enquote {\bibinfo {title} {Hierarchical
  clustering via joint between-within distances: Extending {W}ard's minimum
  variance method},}\ }\href {\doibase 10.1007/s00357-005-0012-9} {\bibfield
  {journal} {\bibinfo  {journal} {Journal of Classification}\ }\textbf
  {\bibinfo {volume} {22}},\ \bibinfo {pages} {151--183} (\bibinfo {year}
  {2005})}\BibitemShut {NoStop}%
\bibitem [{\citenamefont {Wang}\ and\ \citenamefont {Song}(2011)}]{Wang2011}%
  \BibitemOpen
  \bibfield  {author} {\bibinfo {author} {\bibfnamefont {H.}~\bibnamefont
  {Wang}}\ and\ \bibinfo {author} {\bibfnamefont {M.}~\bibnamefont {Song}},\
  }\bibfield  {title} {\enquote {\bibinfo {title} {Ckmeans.1d.dp: Optimal
  k-means clustering in one dimension by dynamic programming},}\ }\href
  {\doibase 10.32614/rj-2011-015} {\bibfield  {journal} {\bibinfo  {journal}
  {The {R} Journal}\ }\textbf {\bibinfo {volume} {3}},\ \bibinfo {pages}
  {29--33} (\bibinfo {year} {2011})}\BibitemShut {NoStop}%
\bibitem [{\citenamefont {Zhou}\ \emph {et~al.}(2020)\citenamefont {Zhou} \emph
  {et~al.}}]{Lancet}%
  \BibitemOpen
  \bibfield  {author} {\bibinfo {author} {\bibfnamefont {F.}~\bibnamefont
  {Zhou}} \emph {et~al.},\ }\bibfield  {title} {\enquote {\bibinfo {title}
  {Clinical course and risk factors for mortality of adult inpatients with
  {COVID}-19 in {W}uhan, {C}hina: a retrospective cohort study},}\ }\href
  {\doibase 10.1016/s0140-6736(20)30566-3} {\bibfield  {journal} {\bibinfo
  {journal} {The Lancet}\ }\textbf {\bibinfo {volume} {395}},\ \bibinfo {pages}
  {1054--1062} (\bibinfo {year} {2020})}\BibitemShut {NoStop}%
\bibitem [{\citenamefont {Mayberry}, \citenamefont {Najjar},\ and\
  \citenamefont {Pietromarchi}(2020)}]{al_jazeera_2020}%
  \BibitemOpen
  \bibfield  {author} {\bibinfo {author} {\bibfnamefont {K.}~\bibnamefont
  {Mayberry}}, \bibinfo {author} {\bibfnamefont {F.}~\bibnamefont {Najjar}}, \
  and\ \bibinfo {author} {\bibfnamefont {V.}~\bibnamefont {Pietromarchi}},\
  }\href@noop {} {\enquote {\bibinfo {title} {Italy coronavirus death toll to
  107, 3089 cases: Live updates},}\ }\bibinfo {howpublished}
  {\url{https://www.aljazeera.com/news/2020/03/italy-death-toll-jumps-global-outbreak-deepens-live-updates-200303233420584.html}}
  (\bibinfo {year} {2020}),\ \bibinfo {note} {{Al Jazeera}, Accessed March 5,
  2020}\BibitemShut {NoStop}%
\bibitem [{\citenamefont {Kantis}, \citenamefont {Kiernan},\ and\ \citenamefont
  {Bardi}(2020)}]{global_health_2020}%
  \BibitemOpen
  \bibfield  {author} {\bibinfo {author} {\bibfnamefont {C.}~\bibnamefont
  {Kantis}}, \bibinfo {author} {\bibfnamefont {S.}~\bibnamefont {Kiernan}}, \
  and\ \bibinfo {author} {\bibfnamefont {J.}~\bibnamefont {Bardi}},\
  }\href@noop {} {\enquote {\bibinfo {title} {Updated: Timeline of the
  {C}oronavirus},}\ }\bibinfo {howpublished}
  {\url{https://www.thinkglobalhealth.org/article/updated-timeline-coronavirus}}
  (\bibinfo {year} {2020}),\ \bibinfo {note} {{Think Global Health}, Accessed
  April 25, 2020}\BibitemShut {NoStop}%
\bibitem [{\citenamefont {McCurry}(2020)}]{guardian_2020}%
  \BibitemOpen
  \bibfield  {author} {\bibinfo {author} {\bibfnamefont {J.}~\bibnamefont
  {McCurry}},\ }\href@noop {} {\enquote {\bibinfo {title} {Test, trace,
  contain: how {S}outh {K}orea flattened its coronavirus curve},}\ }\bibinfo
  {howpublished}
  {\url{https://www.theguardian.com/world/2020/apr/23/test-trace-contain-how-south-korea-flattened-its-coronavirus-curve}}
  (\bibinfo {year} {2020}),\ \bibinfo {note} {{The Guardian}, Accessed April
  23, 2020}\BibitemShut {NoStop}%
\bibitem [{\citenamefont {McDonell}(2020)}]{bbc_2020}%
  \BibitemOpen
  \bibfield  {author} {\bibinfo {author} {\bibfnamefont {S.}~\bibnamefont
  {McDonell}},\ }\href@noop {} {\enquote {\bibinfo {title} {Coronavirus: {US}
  and {A}ustralia close borders to {C}hinese arrivals},}\ }\bibinfo
  {howpublished} {\url{https://www.bbc.com/news/world-51338899}} (\bibinfo
  {year} {2020}),\ \bibinfo {note} {{BBC}, Accessed February 2,
  2020}\BibitemShut {NoStop}%
\bibitem [{\citenamefont {McCann}, \citenamefont {Popovich},\ and\
  \citenamefont {Wu}(2020)}]{nyt2020}%
  \BibitemOpen
  \bibfield  {author} {\bibinfo {author} {\bibfnamefont {A.}~\bibnamefont
  {McCann}}, \bibinfo {author} {\bibfnamefont {N.}~\bibnamefont {Popovich}}, \
  and\ \bibinfo {author} {\bibfnamefont {J.}~\bibnamefont {Wu}},\ }\href@noop
  {} {\enquote {\bibinfo {title} {Italy’s virus shutdown came too late. what
  happens now?}}\ }\bibinfo {howpublished}
  {\url{https://www.nytimes.com/interactive/2020/04/05/world/europe/italy-coronavirus-lockdown-reopen.html}}
  (\bibinfo {year} {2020}),\ \bibinfo {note} {{The New York Times}, Accessed
  April 5, 2020}\BibitemShut {NoStop}%
\bibitem [{\citenamefont {Matthews}(2020)}]{fp_2020}%
  \BibitemOpen
  \bibfield  {author} {\bibinfo {author} {\bibfnamefont {O.}~\bibnamefont
  {Matthews}},\ }\href@noop {} {\enquote {\bibinfo {title} {Britain drops its
  go-it-alone approach to coronavirus},}\ }\bibinfo {howpublished}
  {\url{https://foreignpolicy.com/2020/03/17/britain-uk-coronavirus-response-johnson-drops-go-it-alone/}}
  (\bibinfo {year} {2020}),\ \bibinfo {note} {{Foreign Policy}, Accessed March
  19, 2020}\BibitemShut {NoStop}%
\bibitem [{wor(2020)}]{worldindata2020}%
  \BibitemOpen
  \href@noop {} {\enquote {\bibinfo {title} {Our {W}orld in {D}ata},}\
  }\bibinfo {howpublished}
  {\url{https://ourworldindata.org/coronavirus-source-data}} (\bibinfo {year}
  {2020}),\ \bibinfo {note} {accessed: 2020-04-30}\BibitemShut {NoStop}%
\bibitem [{\citenamefont {Milligan}(1980)}]{Milligan1980}%
  \BibitemOpen
  \bibfield  {author} {\bibinfo {author} {\bibfnamefont {G.~W.}\ \bibnamefont
  {Milligan}},\ }\bibfield  {title} {\enquote {\bibinfo {title} {An examination
  of the effect of six types of error perturbation on fifteen clustering
  algorithms},}\ }\href {\doibase 10.1007/bf02293907} {\bibfield  {journal}
  {\bibinfo  {journal} {Psychometrika}\ }\textbf {\bibinfo {volume} {45}},\
  \bibinfo {pages} {325--342} (\bibinfo {year} {1980})}\BibitemShut {NoStop}%
\bibitem [{\citenamefont {Rousseeuw}(1987)}]{Rousseeuw1987}%
  \BibitemOpen
  \bibfield  {author} {\bibinfo {author} {\bibfnamefont {P.~J.}\ \bibnamefont
  {Rousseeuw}},\ }\bibfield  {title} {\enquote {\bibinfo {title} {Silhouettes:
  A graphical aid to the interpretation and validation of cluster analysis},}\
  }\href {\doibase 10.1016/0377-0427(87)90125-7} {\bibfield  {journal}
  {\bibinfo  {journal} {Journal of Computational and Applied Mathematics}\
  }\textbf {\bibinfo {volume} {20}},\ \bibinfo {pages} {53--65} (\bibinfo
  {year} {1987})}\BibitemShut {NoStop}%
\bibitem [{\citenamefont {Krzanowski}\ and\ \citenamefont
  {Lai}(1988)}]{krzanowski1988}%
  \BibitemOpen
  \bibfield  {author} {\bibinfo {author} {\bibfnamefont {W.~J.}\ \bibnamefont
  {Krzanowski}}\ and\ \bibinfo {author} {\bibfnamefont {Y.~T.}\ \bibnamefont
  {Lai}},\ }\bibfield  {title} {\enquote {\bibinfo {title} {A criterion for
  determining the number of groups in a data set using sum-of-squares
  clustering},}\ }\href {\doibase 10.2307/2531893} {\bibfield  {journal}
  {\bibinfo  {journal} {Biometrics}\ }\textbf {\bibinfo {volume} {44}},\
  \bibinfo {pages} {23--34} (\bibinfo {year} {1988})}\BibitemShut {NoStop}%
\bibitem [{\citenamefont {Hubert}\ and\ \citenamefont
  {Levin}(1976)}]{Hubert1976}%
  \BibitemOpen
  \bibfield  {author} {\bibinfo {author} {\bibfnamefont {L.~J.}\ \bibnamefont
  {Hubert}}\ and\ \bibinfo {author} {\bibfnamefont {J.~R.}\ \bibnamefont
  {Levin}},\ }\bibfield  {title} {\enquote {\bibinfo {title} {A general
  statistical framework for assessing categorical clustering in free recall.}}\
  }\href {\doibase 10.1037/0033-2909.83.6.1072} {\bibfield  {journal} {\bibinfo
   {journal} {Psychological Bulletin}\ }\textbf {\bibinfo {volume} {83}},\
  \bibinfo {pages} {1072--1080} (\bibinfo {year} {1976})}\BibitemShut {NoStop}%
\bibitem [{\citenamefont {McClain}\ and\ \citenamefont
  {Rao}(1975)}]{Mcclain1975}%
  \BibitemOpen
  \bibfield  {author} {\bibinfo {author} {\bibfnamefont {J.~O.}\ \bibnamefont
  {McClain}}\ and\ \bibinfo {author} {\bibfnamefont {V.~R.}\ \bibnamefont
  {Rao}},\ }\bibfield  {title} {\enquote {\bibinfo {title} {{CLUSTISZ}: A
  program to test for the quality of clustering of a set of objects},}\
  }\href@noop {} {\bibfield  {journal} {\bibinfo  {journal} {Journal of
  Marketing Research}\ }\textbf {\bibinfo {volume} {12}},\ \bibinfo {pages}
  {456--460} (\bibinfo {year} {1975})}\BibitemShut {NoStop}%
\bibitem [{\citenamefont {Dunn}(1974)}]{Dunn1974}%
  \BibitemOpen
  \bibfield  {author} {\bibinfo {author} {\bibfnamefont {J.~C.}\ \bibnamefont
  {Dunn}},\ }\bibfield  {title} {\enquote {\bibinfo {title} {Well-separated
  clusters and optimal fuzzy partitions},}\ }\href {\doibase
  10.1080/01969727408546059} {\bibfield  {journal} {\bibinfo  {journal}
  {Journal of Cybernetics}\ }\textbf {\bibinfo {volume} {4}},\ \bibinfo {pages}
  {95--104} (\bibinfo {year} {1974})}\BibitemShut {NoStop}%
\end{thebibliography}%
\end{document}